\documentclass[12pt,aps]{revtex4}
\usepackage[cp1251]{inputenc}

\begin{document}
\input{epsf}

\title{On Electromagnetically Induced Transparency
in the Degenerate $\Lambda$-scheme}

\author{V.S. Zapasskii}
\begin{abstract}
Saint-Petersburg State University Physics Department,
Saint-Petersburg, 198504,  Russia
\vskip10mm

Using an experimental work on “stopped light” as an example, we
show how a classical phenomenon of linear optics – interference of
polarized light – can imitate  the effect of electromagnetically
induced transparency
\end{abstract}

\maketitle

 After the first impressive experiments on
the so-called “slow light” \cite{kasapi,hau}, there have been
reported observations, no less spectacular, made under much
simpler experimental conditions. In one of the most popular
publication of this sort \cite{lukin}, the symptoms of retarded
polarization-optical response of an atomic medium were regarded as
manifestations of the effects of “slow” and “stopped” light (for
more detail, see \cite{alex}). In this note, we attract attention
to an instructive aspect of this misinterpretation, which, in
fact, resulted from erroneous identification of the interference
of polarized beams as the effect of electromagnetically induced
transparency (EIT).

    Recall that the EIT effect \cite{har} is revealed, in its simplest form,
    under the action of a sufficiently strong resonant field onto a three-level quantum
    system with the $\Lambda$-type energy diagram (Fig. 1). It appears that the action of
    the pump (control) field $E_c$ in one arm of the $\Lambda$-scheme (with both the initial and final
    levels of the appropriate transition being empty) renders the system transparent for
    the weak probe (signal) field $E_s$ acting in the second arm provided that the frequency
    difference between the two fields coincides with the frequency of transition between the two lowest levels of the $\Lambda$-scheme.

The narrow dip formed under these conditions in the absorption
spectrum of the system (in the channel of the probe light
$|1\rangle$ - $|3\rangle$, Fig. 1) provides high steepness of the
refractive index dispersion in the vicinity of the transparency
point and can be revealed, in particular, in a giant reduction of
the group velocity of light propagating in this medium.  This is a
gist of the slow light effect based on the phenomenon of
electromagnetically induced transparency.

Turning to physical content of the EIT, we can recall that this
effect is, in essence, a strongly asymmetric (with respect to
intensities of the two fields) version of the effect of coherent
population trapping \cite{agap} usually detected as a dip in the
luminescence excitation spectrum under comparable intensities of
the fields. In its turn, the effect of coherent population
trapping is a direct development of the effect of optically driven
spin precession \cite{bell}, with magnetic sublevels of the ground
state serving as the lowest levels of the $\Lambda$-scheme. And,
eventually, the effect of optically driven spin precession is
nothing else than the effect of optical orientation (or optical
alignment) in the rotating coordinate frame \cite{happer}. So,
without restraining specificity of the EIT --- the only inear
(with respect to the probe light) effect among all the above
resonant phenomena, we can say that EIT is just a modification of
the effect of generalized optical alignment. In [3], the attempt
was made to realize the EIT effect, so to say, on the lowest step
of the above hierarchy, when the lowest levels of the
$\Lambda$-scheme are the degenerate magnetic sublevels of the
ground state.

The experiments of [3] were performed on transition $5^2S_{1/2}, F
= 2 - 5^2P_{1/2}, F = 1$ of $^{87}$Rb using a tunable diode laser
as a light source and a Pockels cell for controlling the beam
polarization. A specific feature of the experiment was that in the
capacity of the control and signal (probe) beams were used
orthogonal polarization components of the same laser beam. In
other words, real variations of the light polarization were
produced using the Pockels cell and were considered as a result of
admixture of the orthogonally polarized probe beam to the control
one. A simplified schematic of the experiment is shown in Fig. 2.
The strong (control) and weak (signal) beams are combined on the
input polarizing beamsplitter and then, after passing through the
atomic system, are split again using the same polarization basis
and are registered by photodetectors.

The basic experimental fact reported by the authors of [3], which,
in authors’ opinion, should serve as the direct evidence for the
EIT effect, is that the system under study is opaque for the probe
light in the absence of the control beam and becomes virtually
completely transparent in its presence. The authors consider this
effect as that of nonlinear optics, the effect of optical
controlling transmission of the medium. At first glance, it looks
convincing. But is it really the case? Is it necessary that the
optical element placed into the “black box” contoured in Fig. 2 by
dashed line should be nonlinear to demonstrate the above effect of
“controlling light by light”?  Let us show that it is not.

We will assume that the control and probe beams are polarized
linearly rather than circularly (as in [3]). This changes
absolutely nothing from the viewpoint of physics (because the
system under study is amenable both to orientation and alignment),
but makes the treatment more visual. We will also assume that the
mutually coherent fields are in-phase, so that being summed at the
input beamsplitter they form a linearly polarized light. The angle
$\varphi$ between the plane of polarization of this light and that
of the probe beam will evidently be equal to arctan$(E_c/E_s)$,
where $E_c$ and $E_s$ are the amplitudes of the control and probe
waves (Fig. 3).

Let us replace the cell with rubidium vapor in the “black box” by
a linear polarizer with its polarizing direction making the same
angle $\varphi$ with the polarization plane of the probe. As a
result, in the absence of the control field, the probe beam, on
its way to the detector, will be first attenuated by this
polarizer (according to the Malus law, the attenuation factor is
$\cos^2\varphi$) and then, in the same degree, by the output
beamsplitter (Fig. 3a). It is clear that total attenuation of the
probe beam determined by the factor $\cos^4\varphi$ can be
formally arbitrary large provided that the angle $\varphi$ is
close enough to 90$^0$.

So, transmission of the probe beam by the “black box” is rather
low. Let us turn the control light on. Now the light outcoming
from the input beamsplitter is again linearly polarized but its
polarization plane exactly coincides with the polarizing direction
of the polarizer in the “black box”, and the light passes through
the polarizer without any attenuation and any change in its
polarization state. In other words, with the control beam turned
on, the “black box” becomes transparent for the whole incident
light and, in particular, --- what is important for us - for its
horizontal polarization component, i.e., for the probe beam.

This is practically exact copy of the EIT effect described in
\cite{lukin} (neglecting dynamics of the process). In reality, the
medium studied in [3] is nonlinear; the pump beam aligns it
optically and thus controls its anisotropy. It means that the
azimuth of the transmission axis of the “black box” is being set
not manually, as in our hypothetical experiment, but rather by the
light wave. For this reason, there arise some apparent
discrepancies between the picture described above and observations
of \cite{lukin}.  For instance, when, in our simplified
arrangement, the control light is on, the probe light can be
detected  at the exit of the scheme even being absent at the
entrance. However, as one can easily see, this is exactly what is
observed in [3] in the demonstrations of “stopped light”, when,
after preliminary manipulations with polarization of the light
beam and subsequent dark pause, the signal in the channel of the
probe light is observed with no probe light at the entrance. In
the experiment, this signal cannot be observed for a long time
because, after a certain time interval, the control beam aligns
the atomic system along its polarization plane or, in other words,
sets properly the polarizer of the “black box” ($\varphi$ = 90
$^o$, Fig. 3), and the projection of the control beam onto the
probe light channel vanishes. In our model, position of the “black
box” polarizer is fixed, and, for this reason, the signal of the
probe light does not disappear.

Thus, we see how easy the EIT effect can be simulated by
interference of polarized light when the “control” and “probe”
fields are just two orthogonal polarization components of the same
beam.

There arises, however, a question: Perhaps this is not am
imitation of the EIT effect but rather the EIT effect proper or,
better to say, what it turns into in the degenerate
$\Lambda$-scheme?  To a certain extent, this is true. But, first
of all, there are no grounds to rename old, well known effects of
classical optics unless we discovered anything new. Second, and
most important, is that “freezing” phase difference between the
probe and control waves in the degenerate $\Lambda$-scheme changes
{\it symmetry} of the problem. In the standard nondegenerate case
of the EIT, anisotropy of the medium is provided exclusively by
the control light, while the weak probe light monitors it in a
nonperturbing way. In the case of the degenerate $\Lambda$-scheme
considered here, the probe light of arbitrarily low intensity
affects anisotropy of the medium (in what way, depends on the
value of the “frozen” phase) and, for this reason, is
fundamentally unable to fulfill its function.   In addition,
optical medium with the anisotropy axis O
 - O$^{prime}$
 cannot be characterized by any optical constant with respect to
probe light polarized along the $x$-axis, and, correspondingly,
one cannot ascribe any phase or group velocity to this light
\cite{alex,kozlov}. In fact, the standard pattern of the EIT
effect is destroyed when the inverse frequency spacing between the
levels $|1\rangle$ and $|2\rangle$ becomes comparable with the
time of measurement.

Note, in conclusion, that the apparent effect of controlling light
by light in a single interference order is a trivial phenomenon.
As applied to the polarization geometry under consideration, it
can be described, in a comprehensive way, by the standard
operation of multiplying Jones vector of the input light by the
matrix of linear polarizer. Still, in some experimental
situations, recognition of this phenomenon appears to be not so
trivial. At least, paper [3], in which exactly this type of error
was made, has gained unquestioning recognition of scientific
community and remains so far widely cited in publications on “slow
light”. This is why we considered this story to be instructive and
deserving additional attention.

The author is grateful to E.B.Aleksandrov for useful discussions.

Fig. 1. Energy-level diagram ($\Lambda$-scheme) illustrating
common approach  to observation of the coherence resonance of the
states $|1\rangle$  and $|2\rangle$.  Specific feature of the EIT
effect is that it is observed at $E_c$ $>>$ $E_s$ and is linear
with respect to probe field $E_s$, while the effects of coherent
population trapping and optically driven spin precession are
observed in the fields of comparable amplitudes (more correctly,
of comparable Rabi frequencies at the transitions $|1\rangle$ -
$|3\rangle$ and $|2\rangle$ - $|3\rangle$).

Fig. 2. Simplified schematic of the experimental setup for
observation of the EIT effect in the degenerate $\Lambda$-scheme.
$E_c$ and $E_s$ are the control and probe (signal) beams, BS –
polarizing beamsplitters, PD – photodetectors, and O  –  object
under study.

Fig. 3. Vector diagram describing transformation of the probe wave
field $E_s$ in the arrangement of Fig. 2 with a polarizer in the
“black box”  in the absence (a) and in the presence (b) of the
control field $E_c$. The fields $E_s$ and $E_c$ are polarized
along the $x$- and $y$-axes, respectively. The line O –
O$^{\prime}$ is directed along the polarizing direction of the
polarizer. ${\bf E_s}^{in}$ and ${\bf E_s}^{out}$ are the probe
field vectors at the entrance and at the exit of the optical
system.

\begin{figure}
\epsfxsize=400pt
\epsffile{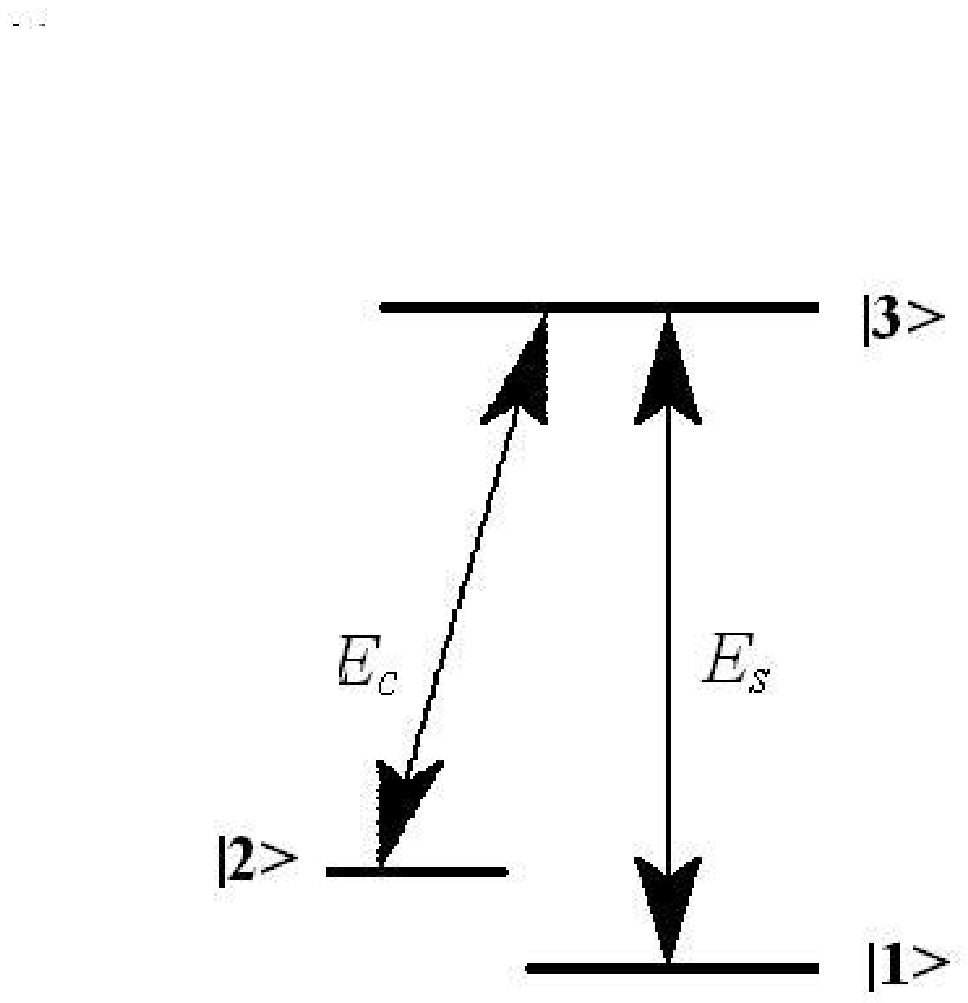}
\caption{Energy-level diagram
($\Lambda$-scheme) illustrating common approach  to observation of
the coherence resonance of the states $|1\rangle$  and
$|2\rangle$.  Specific feature of the EIT effect is that it is
observed at $E_c$ $>>$ $E_s$ and is linear with respect to probe
field $E_s$, while the effects of coherent population trapping and
optically driven spin precession are observed in the fields of
comparable amplitudes (more correctly, of comparable Rabi
frequencies at the transitions $|1\rangle$ - $|3\rangle$ and
$|2\rangle$ - $|3\rangle$).}
\end{figure}

\begin{figure}
\epsfxsize=400pt \epsffile{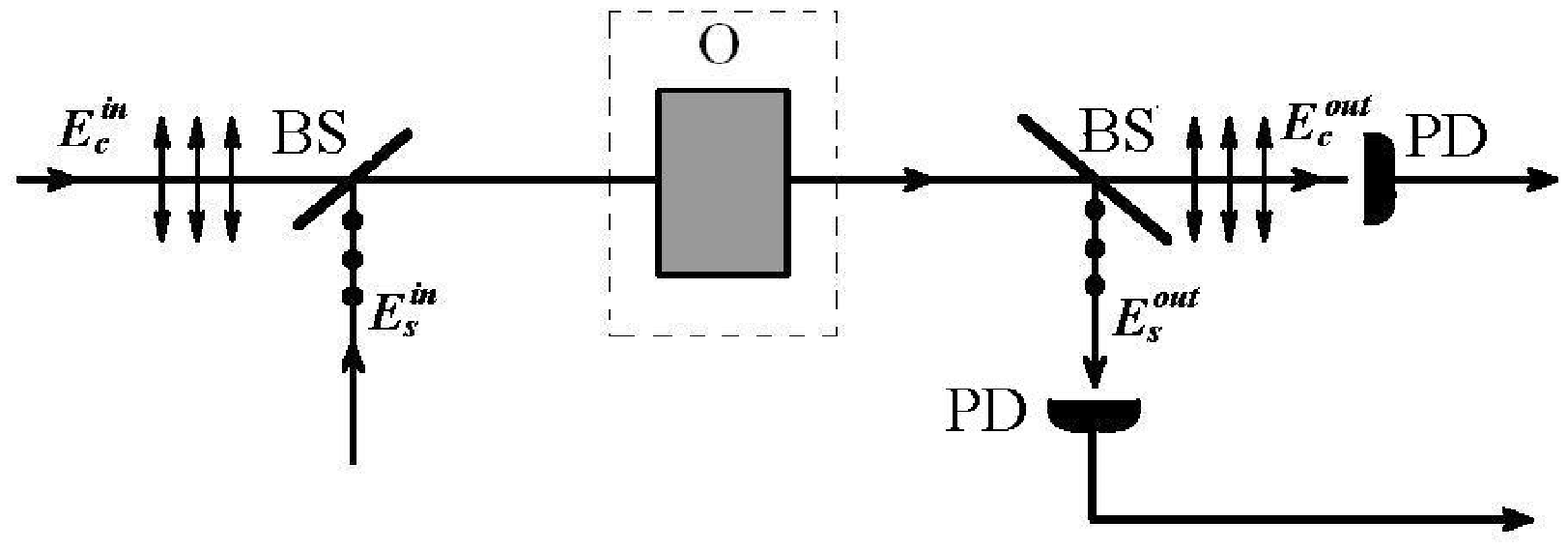} \caption{Simplified schematic
of the experimental setup for observation of the EIT effect in the
degenerate $\Lambda$-scheme. $E_c$ and $E_s$ are the control and
probe (signal) beams, BS – polarizing beamsplitters, PD –
photodetectors, and O  –  object under study.}
\end{figure}

\begin{figure}
\epsfxsize=400pt \epsffile{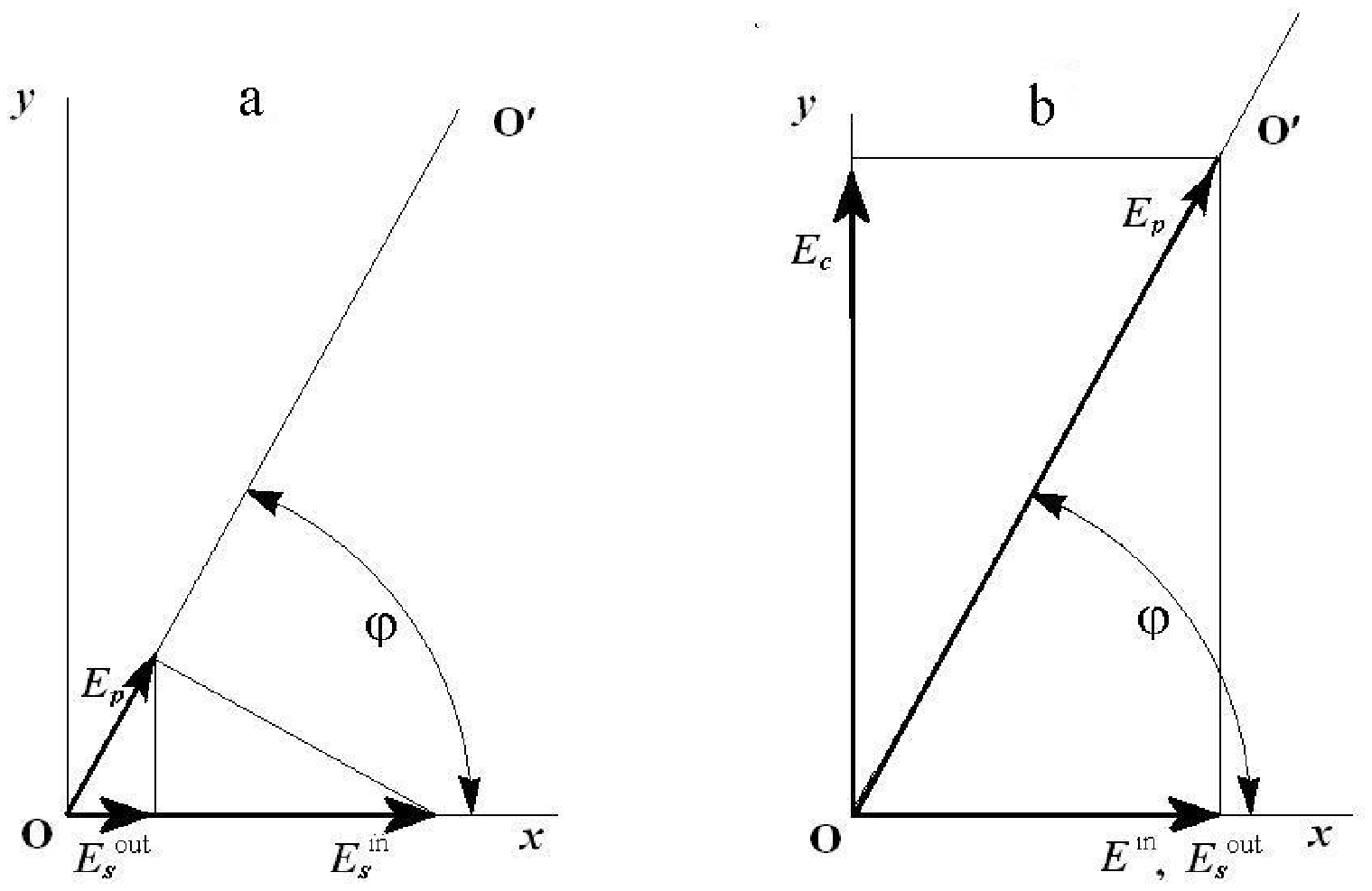} \caption{Vector diagram
describing transformation of the probe wave field $E_s$ in the
arrangement of Fig. 2 with a polarizer in the “black box”  in the
absence (a) and in the presence (b) of the control field $E_c$.
The fields $E_s$ and $E_c$ are polarized along the $x$- and
$y$-axes, respectively. The line O – O$^{\prime}$ is directed
along the polarizing direction of the polarizer. ${\bf E_s}^{in}$
and ${\bf E_s}^{out}$ are the probe field vectors at the entrance
and at the exit of the optical system.}
\end{figure}

\end{document}